\documentclass[a4paper,12pt]{article}
\usepackage{amsfonts}
\usepackage{latexsym}
\usepackage{amsmath}
\usepackage{amssymb}
\usepackage{amssymb}
\usepackage{slashed}
\usepackage{upgreek}
\usepackage{mathrsfs}
\usepackage{bbm}

\usepackage{amsthm}

\theoremstyle{remark}

\usepackage[utf8]{inputenc}
\usepackage[english]{babel}

\theoremstyle{definition}

\allowdisplaybreaks

\usepackage{relsize}
\usepackage{graphicx}
\usepackage{comment}

\renewcommand{\thefootnote}{\fnsymbol{footnote}}

\def\appendix#1{\addtocounter{section}{1}\setcounter{equation}{0}
\renewcommand{\thesection}{\Alph{section}}
\section*{Appendix \thesection\protect\indent \parbox[t]{11.15cm}{#1}}
\addcontentsline{toc}{section}{Appendix \thesection\ \ \ #1}}

\newcommand{\bea}{\begin{eqnarray}}
\newcommand{\eea}{\end{eqnarray}}

\usepackage[usenames,dvipsnames,svgnames,table]{xcolor}	% AH -- REMOVE BEFORE UPLOAD
\usepackage[normalem]{ulem}				% AH -- REMOVE BEFORE UPLOAD
\usepackage{microtype}
\usepackage[colorlinks, citecolor=blue, linkcolor=blue, urlcolor=Maroon, filecolor=Maroon, linktocpage=true]{hyperref} % AH

\usepackage{chngcntr}
\counterwithout{equation}{section}

\begin{document}

\begin{center}
%\today
\vspace*{-1.0cm}
\begin{flushright}
%\normalsize{\texttt{ZMP-HH/17-24}}\\
\end{flushright}
%\hfill hep-th/yymmnnn \\
%\hfill UB-ECM-PF-06-43 \\
%\hfill DMUS--MP--13/06 \\

%\vspace{2.0cm} {\Large \bf Covariant form hierarchies, integrability  and Killing spinor equations } \\[.2cm]

\vspace{2.0cm} {\Large \bf Separability, plane wave limits and  black holes } \\[.2cm]

\vskip 2cm
 G.  Papadopoulos
\\
\vskip .6cm

%\begin{small}
%${}^1$ \textit{Department of Mathematics, King's College London
%\\
%Strand, London WC2R 2LS, UK}\\
%\texttt{sebastian.lautz@kcl.ac.uk}\\
%\end{small}
%\vskip0.5cm

\begin{small}
\textit{Department of Mathematics
\\
King's College London
\\
Strand
\\
 London WC2R 2LS, UK}\\
\texttt{george.papadopoulos@kcl.ac.uk}
\end{small}
\\*[.6cm]

\end{center}

\vskip 2.5 cm

\begin{abstract}
\noindent
We present a systematic construction  of the Penrose coordinates and plane wave limits of spacetimes for which both the null Hamilton-Jacobi and geodesic equations separate. The  method  is illustrated for  the Kerr-NUT-(A)dS four-dimensional black holes. The plane wave limits of   the near horizon geometry of the extreme Kerr black hole are also explored. All near horizon geometries of extreme black holes with a Killing horizon admit Minkowski spacetime as  a plane wave limit.

\end{abstract}

%\vskip 1cm

%{\small Keywords: spacetime geometry; black holes; special Lorentzian structures; G-structures}

\newpage

\renewcommand{\thefootnote}{\arabic{footnote}}
%\tableofcontents

%%%%%%%%%%%%%%%%%%%%%%%%%%%%%%%%%%%%%%%%%%%%%%%%%%%%%%%%%%%%%%%%%%%%%%%%%%
%\setcounter{section}{0}\setcounter{equation}{0}
%%%%%%%%%%%%%%%%%%%%%%%%%%%%%%%%%%%%%%%%%%%%%%%%%%%%%%%%%%%%%%%%%%%%%%%%%%

%\section{Introduction}
%%\counterwithout{equation}{chapter}

\section{Introduction}

It has been known for sometime \cite{carter-a, carter-b} that geodesic equation of many 4- and higher-dimensional black holes can be separated, see also the comprehensive reviews \cite{revky, frolov} and references within. This means that there is a set of coordinates on the spacetime such that  sequentially the geodesic equation can be expressed as a set of ordinary differential equations each depending on a single unknown function. In principle these equations can be solved\footnote{Separability does not necessarily imply integrability provided that the latter means that the solutions could  be expressed in terms of some ``simple'' functions.} to find all the geodesics of the spacetime. The separability of the geodesic equation is due to the existence of ``hidden'' constants of motion in addition to those associated with the  (commuting) isometries of the spacetime metric \cite{penrose-a, penrose-b, floyd}.    This extends to other differential equations on the spacetime, like the Hamilton-Jacobi (HJ), Klein-Gordon and Dirac equations \cite{carter-b, chandrasekhar}.

Although, the problem of finding the orbits of test particles in such a spacetime has in principle been resolved in the way explained above, the problem remains in the context of string theory. It is well known that the equations of motion for strings  have only been solved on spacetimes that exhibit a large group  of symmetries, like Minkowski spacetime, group manifolds, some homogeneous spaces and some of their discrete quotients, see e.g. \cite{witten, polchinski} and references within. The list does not include 4-dimensional black holes and cosmological spacetimes.
As a result it is not known how strings behave for example  near the black hole and cosmological singularities.  Though significant progress has been made towards the understanding the thermodynamical properties of (extreme) black holes using holography and AdS/CFT techniques, see e.g. \cite{strominger-a, strominger-b, maldacena}.

To get a glimpse of how strings behave in black hole and cosmological spacetimes, it has been advocated in \cite{gp-a, gp-b} to explore the propagation of strings at the plane wave  limits of these spacetimes. Sometime ago Penrose has shown  that on every spacetime one can define  a  set of coordinates adapted to a null geodesic congruence\footnote{We shall refer to these as Penrose coordinates.} which in turn can be used to construct a plane wave or Penrose limit \cite{penrose-c} for the spacetime.
The plane wave limits that probe the singularities of  a large class of spherically symmetric black holes, that include the Schwarzschild and Reissen-Nordstr\"om,  have been explored in \cite{gp-a, gp-b, gp-x} and has been found that the plane wave
profile in Brinkmann coordinates is $A\sim \lambda^{-2}$, where at $\lambda=0$ is the black hole singularity. A similar plane wave profile is also exhibited by cosmological singularities. Furthermore in \cite{gp-c} it was found that that there is a prescription such that string massive modes can propagate through the singularity from the region $\lambda>0$ to the $\lambda<0$ region while particles do not.  This indicates that strings exhibit a ``softer'' behaviour near singularities than particles.  Conceivably the Penrose diagrams of spacetimes probed with test strings may look different from the standard ones that exhibit the behaviour of test particles, i.e. that of geodesics.
Other applications of plane wave limits are those of the AdS near horizon geometries of brane and black hole spacetimes \cite{gp-d}. These plane wave limits \cite{gp-e} have found extensive applications in string theory \cite{metsaev} and in AdS/CFT \cite{maldacena-b}, see also \cite{mattias} for a more recent work.

As there is no method so far  which can be utilized to solve  string theory on black hole spacetimes, one way to proceed is to focus on the plane wave limits and proceed to study string theory at those limits. As the null geodesic and HJ equations are both separable for a large selection of black-hole spacetimes, the plane wave limits can be constructed in a systematic way. The purpose of this paper is first to give a general description of the method that can be used to construct the Penrose coordinates (PCs) on a spacetime whenever both
the geodesic and HJ equations are separable. Using these, the associated plane wave limits are also presented.  We illustrate the method with the description of the PCs and plane wave limits of  Kerr-NUT-(A)dS black holes. Then we explore some examples with focus on the near horizon geometry of the Kerr black hole. This is the analogue of the AdS near horizon geometries of some supersymmetric black holes and branes but now in a non-supersymmetric setting. It turns out that every near horizon geometry of an extreme Killing horizon has a plane wave limit which is the Minkowski spacetime.

This paper is organised as follows. In section 2, we describe the construction of PCs for spacetimes with separable HJ and geodesic equations. In section 3, we apply the formalism that has been developed first to Kerr-NUT-(A)dS black holes and then to the near horizon geometry of the extreme Kerr black hole, and in section 4 we give our conclusions.

 %As it will become apparent later, spacetimes can admit many different Penrose limits which may depend on the different null twist free geodesic congruences that one can construct on a spacetime. All null geodesics in the same twist free null geodesic congruence give the same Penrose limit.

\section{Separability of geodesic systems and Penrose coordinates}

\subsection{Hamilton-Jacobi function and Penrose coordinates}

A detailed description of the construction of Penrose coordinates on a spacetime associated to a null geodesic can be found in \cite{gp-a, gp-b}. Here for completeness we shall repeat some essential steps in the construction  and make the computation a bit more explicit. To begin consider the null HJ equation
\bea
g^{\mu\nu} \partial_\mu S \partial_\nu S=0~,
\label{nhj}
\eea
where $S$ is the HJ function.
This is a first order partial differential equation and locally there is a unique solution  provided that a boundary condition for $S$ is specified on a spatial hyper-surface $F(x^\mu)=0$ in a spacetime. In practise though such boundary value problems are hard to solve. Instead the aim is to find the ``complete'' solution to the null HJ equation\footnote{However note that if $S$ is a solution to the null HJ equation, then $G(S)$ for any (smooth) function $G$ is also a solution.}. Such a solution  is characterized  by the presence of  $n$  integration constants, where  $n$ is the dimension of the spacetime. Two of the constants are forgetful as if $S$ is a solution, then $ p S+ c$ is also a solution for any constants $p$ and $c$. Thus the complete solutions of the null HJ equation is specified by $n-2$ independent constants $\alpha$.

Next observe that the solutions of
\bea
\dot x^\mu=g^{\mu\nu}\partial_\nu S
\label{geo1}
\eea
are null geodesics, where $\dot x^\mu={dx^\mu\over d\lambda}$ and  $\lambda$ is the affine parameter.
 Indeed $g_{\mu\nu} \dot x^\mu \dot x^\nu=0$ as a consequence of (\ref{nhj}) and
 \bea
 \nabla_\lambda \dot x^\mu=g^{\mu\nu} \dot x^\rho \nabla_\rho \partial_\nu S={1\over2} g^{\mu\nu} \partial_\nu\Big(g^{\rho\sigma} \partial_\rho S\, \partial_\sigma S\Big)=0~.
 \eea
 The  system of first order equations conditions (\ref{geo1}) has a unique  solution $x^\mu(\lambda, x_0^\nu, \alpha)$  for small $\lambda$  subject to a boundary condition at $\lambda=0$, $x^\mu(0, x_0^\nu, \alpha)=x_0^\mu$.  Next consider  the solutions of (\ref{geo1}) for different boundary conditions $x_0^\mu$ and assume that they define a null geodesic congruence on the spacetime such that they intersect a hypersurface ${\cal H}$ only once. In such a case after a possible redefinition of the affine parameter of each geodesic, the boundary conditions for the solutions of (\ref{geo1}) at $\lambda=0$ can be taken to be the points of ${\cal H}$.

Furthermore observe that given a solution of (\ref{geo1}), one has that
\bea
S(x^\mu(\lambda, x_0^\nu, \alpha), \alpha)=S( x_0^\nu, \alpha)~,
\eea
i.e. $S$ does not depend on the affine parameter $\lambda$. Indeed
\bea
\dot S=\dot x^\mu \partial_\mu S= g^{\mu\nu} \partial_\mu S \partial_\nu S=0~.
\eea

A Penrose coordinate system on a spacetime is defined as
\bea
\lambda=\lambda~,~~~\{w^i, u\}=\{x^\mu_0, u\}/\{S(x_0^\mu, \alpha)-u,  H(x^\mu_0)\}~,
\eea
where $H(x_0^\mu)=0$ defines the initial value hypersurface ${\cal H}$ and as indicated $w^i$ are the solutions to the conditions $S(x_0^\mu,\alpha)=u$ and $H(x_0^\mu)=0$.  It is assumed that the hypersurface ${\cal H}$ and the level sets of $S$ intersect transversally. The choice of $H$ is not essential as different choices give rise to a different parameterisation of the null geodesic congruence.
Note however that a different choice of integration constants $\alpha$ may lead to a different choice of PCs on a spacetime. This is because a choice of different set of integration constants $\alpha$ may lead to a different null geodesic congruence.

To write explicitly the change of coordinate transformation consider a solution to  (\ref{geo1}) as $x^\mu(\lambda, x_0^\mu, \alpha)$, take  the differential and express it in the new coordinates as
\bea
dx^\mu&=&\dot x^\mu  d\lambda+{\partial x^\mu\over \partial x_0^\nu} ({\partial x^\nu_0\over \partial u} du+{\partial x^\nu_0\over \partial w^i} dw^i)
\cr
&=&g^{\mu\nu} \partial_\nu S\, d\lambda+{\partial x^\mu\over \partial x_0^\nu} ({\partial x^\nu_0\over \partial u} du+{\partial x^\nu_0\over \partial w^i} dw^i)~,
\eea
where the integration constants $\alpha$ are considered fixed and they will be suppressed from now on.
In these new coordinates\footnote{Similar coordinates can be adapted to time-like ($m^2>0$) and space-like ($m^2<0$) geodesics, $m$ constant. The metric in the new coordinates reads
 as $g=m^2 d\lambda^2+2du \,\Big( \, d\lambda+h_i\, dw^i + {1\over2}\Delta\, du\Big)+ \gamma_{ij}\, dw^i\, dw^j$, where $g^{\mu\nu} \partial_\mu S\partial_\nu S=m^2$.} the metric reads as
\bea
g=2du \,\Big( \, d\lambda+h_i(\lambda, u, w)\, dw^i + {1\over2}\Delta(\lambda,u,w)\, du\Big)+ \gamma_{ij}(\lambda,u,w)\, dw^i\, dw^j~,
\label{prcor}
\eea
where we have used
\bea
g_{uu} =g_{\mu\nu} g^{\mu\rho} \partial_\rho S g^{\nu\sigma} \partial_\sigma S=0~,
\eea
as a consequence of (\ref{nhj}) and
\bea
&&g_{\mu\nu} g^{\mu\rho} \partial_\rho S {\partial x^\nu\over \partial x_0^\sigma}\Big({\partial x^\sigma_0\over \partial u} du+ {\partial x^\sigma_0\over \partial w^i} dw^i\Big) d\lambda
=\partial_\nu S {\partial x^\nu\over \partial x_0^\sigma}\Big({\partial x^\sigma_0\over \partial u} du+ {\partial x^\sigma_0\over \partial w^i} dw^i\Big) d\lambda
\cr
&&
=\partial_{x^\nu_0} S(x_0^\rho) {\partial x^\nu_0\over \partial u} d\lambda\, du+\partial_{x^\nu_0} S(x_0^\rho) {\partial x^\nu_0\over \partial w^i} d\lambda\, dw^i
\cr
&&
= {\partial u\over \partial u}d\lambda\, du+{\partial u\over \partial w^i} dw^i d\lambda=du\, d\lambda~.
\eea
The remaining components of the metric can be easily evaluated leading to (\ref{prcor}).

Adapting a set of PCs to a spacetime as in (\ref{prcor}), the plane wave limit to a spacetime is defined after scaling the spacetime metric
as $g\rightarrow \ell^{-2} g$, performing the coordinate transformation $u\rightarrow \ell^2 u$, $w^i\rightarrow \ell w^i$ and $\lambda\rightarrow \lambda$ and taking the limit $\ell\rightarrow 0$. The metric at the limit takes the form
\bea
g_{pw}=2 du\,d\lambda+ \gamma_{ij}(\lambda) dw^i dw^j~.
\eea
This is a plane-wave metric written in Rosen coordinates.  The limit $\ell\rightarrow 0$ restricts the non-vanishing components of the metric at the null godesic with boundary conditions $u=w^i=0$. So the limit depends on the choice of the null geodesic.

The plane wave metric $g_{pw}$ can be rewritten in Brinkmann coodinates as
\bea
g_{pw}=2 d\lambda\Big( dv+{1\over2} A_{ab}(\lambda) x^a x^b d\lambda\Big)+ \delta_{ab} dx^a dx^b~,
\eea
where $w^i=e^i_a(\lambda) x^a$, $u=v-{1\over2} \gamma_{ij} \partial_\lambda e^i_{(a} e^j_{b)} x^a x^b$, with $\gamma_{ij} e^i_a e^j_b=\delta_{ab}$ and
\bea
A_{ab}=\gamma_{ij} \partial_\lambda e^i_a \partial_\lambda e^j_b-\partial_\lambda\big( \gamma_{ij} \partial_\lambda e^i_{(a} e^j_{b)}\big)~,
\eea
 is the plane wave profile mentioned above.

\subsection{Separability  and Penrose coordinates}

In gravitational systems that both the null  geodesic and HJ equations are separable, there is a simplification in the description of PCs of the spacetime.
The separability of the HJ equation means that there is a coordinate system $\{x^\mu\}$ on the spacetime such that the HJ function $S$ can be written as
\bea
S=\sum_\mu S_{(\mu)}(x^\mu)
\eea
where each function $S_{(\mu)}$ depends only on the coordinate $x^\mu$, and the HJ equation reduces to a first order  ordinary differential equation for each $S_{(\mu)}$.
Such differential equations can be integrated to determine  $S_{(\mu)}$ in terms of $x^\mu$ up to an integration constant $\alpha$.

The integrability of the geodesic equations means that amongst the coordinates $\{x^\mu\}$ that separate the null HJ equation, there is a coordinate, say $x^1\in \{x^\mu\}$, such that the geodesic equation $\dot x^1=g^{1\nu} \partial_\nu S$ is an equation that involves only $x^1$ and  the affine parameter $\lambda$, i.e. it does not depend on the remaining spacetime coordinates. As a result  this can be integrated to express
$x^1$ in terms of $\lambda$ as $x^1=x^1(\lambda, x_0^1)$. Then the remaining geodesic equations can be solved sequentially. This means that if the solution for the first k-coordinates is known, stay $x^\mu=x^\mu(\lambda, x_0^1, \dots, x_0^k)$, $\mu=1,\dots, k$, then after a substitution of these solutions into the geodesic  equation for the $k+1$ coordinate, $x^{k+1}$, will lead to an equation which will involve only the coordinate $x^{k+1}$ and $\lambda$. As a result, it can also be integrated to give $x^{k+1}$ in terms of $\lambda$, and so on.

It is clear that the separability of the geodesic equations will lead to a solution of the geodesic equations which can be arranged, after a possible permutation of the coordinates, to be of the form
\bea
x^\mu=x^\mu(\lambda, x_0^1, \dots, x_0^\mu)~,~~~\mu=1,2,3,\dots, n
\eea
To proceed, it is convenient to take $H(x_0^\nu)=x_0^1=0$.  In such a case, the change of coordinate transformation to the Penrose coordinates reads
\bea
dx^1=\dot x^1 d\lambda~;~~~dx^\mu= \dot x^\mu d\lambda +\sum_{\nu=2}^\mu {\partial x^\mu\over \partial x_0^\nu} dx^\nu_0~,~~~\mu>1~;~~~u=S(x_0^\nu)=
\sum_{\mu>1} S_{(\mu)}(x_0^\mu)~.
\eea
For spacetimes with isometries generated by Killing vectors $\xi$, there is a further simplification. Typically the separability coordinates $\{x^\mu\}$ are adapted to the commuting isometries of the spacetime. If there is such an isometry $\xi$ which is adapted to any coordinate $x^\mu$, $\mu>1$, say for simplicity $x^2$, i.e. $\xi=\partial_{x^2}$, then the HJ function reads as
\bea
S=L_{2}\, x^2+ \sum_{\mu>2} S_{(\mu)}(x^\mu)~,
\eea
where $L_{2}$ is a constant. $L_2$  is one of the integration constants $\alpha$ of the HJ equation. For null geodesic congruences for which $L_{2}\not=0$, the coordinate transformation $u=S(x_0^\nu)$ can be solved as
\bea
x^2_0={1\over L_{2}} \big(u- \sum_{\mu>2} S_{(\mu)}(x_0^\mu)\big)~.
\eea
Thus the coordinates $w^i$ can be identified with $x_0^\mu$ for $\mu>2$.

Note that typically there may be several commuting Killing vector fields $\xi$ on a spacetime each one giving a different solution of the equation $u=S(x_0^\nu)$
as explained above. Choosing different non-vanishing integration constants $\alpha$, i.e. different solutions  to the HJ equation,  gives rise to different null geodesic congruences.  This in turn may lead to different Penrose coordinates adapted to the spacetime.

\section{Rotating black holes}

The PCs and associated limits for spherically symmetric black holes have been extensively investigated in \cite{gp-a, gp-b}. Here we shall focus on rotating black holes as described by the Kerr-NUT-(A)dS family. Later we shall specialise on  the Kerr black hole and in particular on the near horizon geometry of the extreme Kerr black hole.

\subsection{ The Kerr-Nut-(A)dS black hole}

The metric of the Kerr-Nut-(A)dS black hole written in the coordinates that both the  geodesic and HJ equations separate  can be expressed as
\bea
g=-{\Delta_r\over \Sigma} (d\tau+y^2 d\psi)^2+{\Delta_y\over \Sigma} (d\tau-r^2 d\psi)^2+{\Sigma\over \Delta_r} dr^2+{\Sigma\over \Delta_y} dy^2~,
\label{nuts}
\eea
where
\bea
&&\Sigma=r^2+y^2~,~~~\Delta_r=(r^2+a^2) (1-{\Lambda\over 3} r^2)-2Mr~,~~~
\cr
&&\Delta_y=(-y^2+a^2) (1+{\Lambda\over 3} y^2)+2Ny~,
\eea
where $M$ is the mass, $a$ is the angular momentum,  and $N$ is the NUT charge of the black hole. The metric above solves the Einstein equations with a cosmological constant $\Lambda$.
For  dS and AdS black holes the cosmological constant is   $\Lambda>0$ and $\Lambda<0$, respectively.

The parameters of the  Kerr black hole are the mass $M$ and the angular momentum $a$ as $\Lambda=N=0$. We also take $|y|<|a|$, $r>0$ and $-\infty<\tau<+\infty$ and restrict the parameters as $M\geq |a|$.

In these coordinates the HJ equation for null geodesics for the metric (\ref{nuts}) can be separated \cite{plebanski} as
\bea
S=-E\tau+L_\psi \psi+ S_{(r)}(r)+ S_{(y)}(y)~,
\eea
where
\bea
\Delta_r (\partial_r S_{(r)})^2={{\cal X}_r\over \Delta_r}~,~~~\Delta_y (\partial_y S_{(y)})^2={{\cal X}_y\over \Delta_y}~,
\eea
and
\bea
{\cal X}_r=(E r^2-L_\psi)^2-\Delta_r K~,~~~{\cal X}_y=-(E y^2+L_\psi)^2+\Delta_y K~,
\eea
with $K$ a separation constant.  For this, we have used that
\bea
g^{-1}={1\over\Sigma}\big[-\Delta_r^{-1} (r^2\partial_\tau+\partial_\psi)^2+ \Delta_y^{-1} (y^2\partial_\tau-\partial_\psi)^2+\Delta_r \partial_r^2+\Delta_y \partial_y^2\big]~.
\eea
 Observe that $\partial_\tau$ and $\partial_\psi$ are commuting isometries and so their adapted coordinates appear linearly in the HJ function $S$.

For the Kerr black hole, the HJ equation does not have solutions if the separation constant $K=0$ and either $E$ or $L_\psi$ are no-vanishing. As there are no no-trivial solutions for $K=E=L_\psi=0$, we shall take $K\not=0$.

Returning to the Kerr-Nut-(A)dS black hole,  the null geodesic equations read as
\bea
\dot\tau=-E g^{\tau\tau}+L_\psi g^{\tau\psi}~,~~~\dot\psi= g^{\psi\psi} L_\psi-g^{\psi\tau}E~,~~~\dot r=g^{rr}\partial_r S_{(r)}~,~~~\dot y= g^{yy} \partial_y S_{(y)}(y)~.
\eea
The last two equations can be solved to express both $r$ and $y$ coordinates in terms of the affine parameter $\lambda$. One can choose the hypersurface ${\cal H}$  either as
$H=r_0=0$ or as $H=y_0=0$. The two choices are equivalent.  Let us take $H=r_0=0$.  In such a case, the null geodesic equations can be formally  integrated to yield
\bea
\tau=\tau(\lambda, y_0)+\tau_0~,~~~\psi=\psi(\lambda, y_0)+\psi_0~,~~~y=y(\lambda, y_0)~,~~~r=r(\lambda)~.
\eea
As a consequence the change of coordinates transformation is
\bea
&&d\tau=\dot\tau\, d\lambda+\partial_{y_0}\tau\, dy_0+d\tau_0~,~~~d\psi=\dot\psi\, d\lambda+ \partial_{y_0}\psi\, dy_0+d\psi_0~,~~~
\cr
&&dy=\dot y\, d\lambda+\partial_{y_0}y\, dy_0~,
dr=\dot r\, d\lambda~,~~~u=S=-E \tau_0+L_\psi \psi_0+S_{(y_0)}(y_0)~,
\eea
To continue the last condition should be solved to express one of the coordinates $\tau_0, \psi_0$ or $y_0$ in terms of $u$. For example, if either $E\not=0$ or $L_\psi\not=0$, then one can easily express $\tau_0$ or $\psi_0$ in terms of $u$ and the remaining coordinates, respectively.  If both $E, L_\psi\not=0$, the two choices are equivalent. Otherwise, the kinematic regimes are different.
In particular if $E\not=0$, then
\bea
\tau_0={1\over E}\big(-u+L_\psi \psi_0+ S_{(y_0)}(y_0)\big)~.
\eea
Clearly the $w$ coordinates are $y_0, \psi_0$.  Using these, the metric can be put in PCs with components
\bea
&&\Delta={1\over E^2\Sigma} \big(\Delta_y-\Delta_r\big)~,~~~h_{\psi_0}={1\over E\Sigma}\Big(\Delta_r \big({1\over E} L_\psi + y^2\big)-\Delta_y \big({1\over E} L_\psi - r^2\big)\Big)~,
\cr
&& h_{y_0}={1\over E \Sigma}\Big(\Delta_r Q_y -\Delta_y Q_r\Big)~,
\cr
&&
\gamma_{y_0y_0}=-{\Delta_r\over\Sigma} Q_y^2
+{\Delta_y\over\Sigma} Q_r^2+{\Sigma\over \Delta_y} \Big({\partial y\over \partial y_0}\Big)^2~,
\cr
&&
\gamma_{\psi_0\psi_0}=-{\Delta_r\over\Sigma}\big({1\over E} L_\psi + y^2\big)^2+{\Delta_y\over\Sigma}\big({1\over E} L_\psi - r^2\big)^2~,
\cr
&&
\gamma_{\psi_0 y_0}=-{\Delta_r\over\Sigma}Q_y\big({1\over E} L_\psi + y^2\big)+{\Delta_y\over\Sigma} Q_r \big({1\over E} L_\psi -r^2\big)~,
\label{nutpmetr}
\eea
where
\bea
Q_y=({\partial \tau\over \partial y_0}+E^{-1}{\partial S_{(y_0)}(y_0)\over \partial y_0}+ y^2 {\partial \psi\over \partial y_0}\big)~,~~~Q_r=({\partial \tau\over \partial y_0}+E^{-1}{\partial S_{(y_0)}(y_0)\over \partial y_0}- r^2 {\partial \psi\over \partial y_0}\big)~.
\eea
One can continue in the same way for $L_\psi\not=0$ to express $\psi_0$ in terms of $u, \tau_0$ and $y_0$. The computation is straightforward and the result will not be stated here.

 Finally if both $E=L_\psi=0$, then $u=S_{(y_0)}(y_0)$.  $y_0$ can be expressed in terms of $u$ by inverting this equation. The $w$ coordinates in this case are $\tau_0$ and $\psi_0$. For the Kerr black hole in this kinematic regime, one has  that
 \bea
S_{(r_0)}=\int^{r_0} dp {K^{{1\over2}}\over |\Delta_p|}~,~~~ S_{(y_0)}(y_0)=\int^{y_0} dp{K^{{1\over2}}\over a^2-p^2}~.
 \eea
where $K>0$ and $r_-<r<r_+$, i.e. the region is between the two horizons of the Kerr black hole.  In this case one finds that
\bea
&&\Delta=-{\Delta_r\over\Sigma} Q^2_y+{\Delta_y\over\Sigma} Q_r~,~~~h_{\tau_0}=-{\Delta_r\over\Sigma} Q_y+{\Delta_y\over\Sigma} Q_r~,~~~h_{\psi_0}=-{\Delta_r\over\Sigma} y^2 Q_y-{\Delta_y\over\Sigma} r^2 Q_r~,~~~
\cr
&&\gamma=-{\Delta_r\over\Sigma}(d\tau_0+y^2 d\psi_0)^2+{\Delta_y\over\Sigma}
(d\tau_0-r^2 d\psi_0)^2~,
\eea
where now
\bea
Q_y={\partial \tau\over \partial u}+ y^2 {\partial \psi\over \partial u}~,~~~Q_r={\partial \tau\over \partial u}-r^2 {\partial \psi\over \partial u}~.
\eea
The finding of closed expressions for the metric in (\ref{nutpmetr}) depends on explicit solutions of the geodesic equation. We shall explore this further  for the near horizon geometry of the extreme Kerr black hole.

\subsection{Near horizon Kerr geometry}

The metric \cite{bardeen} of the near horizon geometry of the extreme Kerr black hole  is
\bea
g=(1+z^2) (-\rho^2 d\tau^2+\rho^{-2} d\rho^2)+{1+z^2\over 1-z^2} dz^2+{1-z^2\over 1+z^2} (2 \rho d\tau+d\varphi)^2~.
\eea
Next introduce the coordinates
\bea
v=\tau-\rho^{-1}~,~~\chi=\varphi-2d\log \rho~,
\eea
to express the above metric as
\bea
g=(1+z^2) (-\rho^2 dv^2+ 2 dv d\rho)+{1+z^2\over 1-z^2} dz^2+{1-z^2\over 1+z^2} (2 \rho dv+d\chi)^2~.
\eea
In these coordinates the HJ equation is separable.
In particular set $S=Ev+L\chi+R(\rho)+Z(z)$,  $E$ and $L$ constants,  to find that
\bea
\rho^2 \Big({dR\over d\rho}\Big)^2+ (2E-4L\rho) {dR\over d\rho}=-Q^2~,~~~\Big({dZ\over dz}\Big)^2+{(1+z^2)^2\over (1-z^2)^2} L^2={Q^2\over 1-z^2}~,
\label{nhorhj}
\eea
where $Q$ is a separation constant. We have also used that
\bea
g^{-1}={2\over 1+z^2} \partial_v \partial_\rho+{\rho^2\over 1+x^2} \partial_\rho^2-{4\rho \over 1+z^2} \partial_\rho \partial_\chi+{1-z^2\over 1+z^2} \partial_z^2+{1+z^2\over 1-z^2} \partial_\chi^2~.
\eea
Further progress depends on exploring (\ref{nhorhj}) and the geodesic equations.

\subsubsection{Null Gaussian Coordinates}

It is well known that all regular (extreme) Killing horizons can be expressed in null Gaussian coordinates, see \cite{isen, gnull}. For the near horizon Kerr geometry, these are a special case of PCs which are associated with null geodesics with  $Q=L=0$; for a different derivation see \cite{li}.  In this kinematic regime
the HJ equation can  be integrated to yield
\bea
S=Ev+2E \rho^{-1}~.
\eea
Moreover, the associated null geodesic equations can  be solved to find
\bea
z=z_0~,~~~\rho=-{E\over 1+z_0^2} \lambda~,~~~u={2\over E} (1+z_0^2) \lambda^{-1}+ u_0~,~~~\chi=-4 \log \lambda+\chi_0~,
\eea
where $\lambda$ is the affine parameter of the geodesics and no integration constant has been introduced for $\rho$. Adapting PCs to this null geodesic congruence,
one finds that
\bea
&&h=\lambda\Big({4(1-x^2)\over (1+x^2)^2} d\phi- {2x\over 1+x^2} dx\Big)~,~~~\Delta=\lambda^2\Big({3-6x^2-x^4\over a^2 (1+x^2)^3}\Big)~,
\cr
&&
\gamma=a^2 {1+x^2\over 1-x^2} dx^2+ 4 a^2 {1-x^2\over 1+x^2} d\phi^2~,
\eea
with $a^2=1$, where we have set $x=z_0, -2\phi=\chi_0-4\log(1+z_0^2), u=S=E v_0$. A non-trivial $a^2$ factor can be introduced after rescaling the whole metric with $a^2$ and changing coordinates as $\lambda \rightarrow a^2 \lambda$. The metric of the near horizon geometry has been put in Gaussian null coordinates, i.e. it has taken the form
\bea
g= 2 du (d\lambda+\lambda\, \tilde h_i dx^i+{1\over2} \lambda^2\, \tilde \Delta du)+ \gamma_{ij} dx^i dx^j~,
\label{ncoor}
\eea
where $\tilde h$, $\tilde \Delta$ and $\gamma$ do not depend {\sl only} on the coordinate $x$.

It has already been mentioned in the beginning of this section  on all near horizon geometries of extreme regular Killing horizons  one can adapt null gaussian  coordinates. In these coordinates,   the metric takes the form (\ref{ncoor}) and  its components $\tilde h$, $\tilde \Delta$ and $\gamma$  do not depend on the coordinates $u$ and $\lambda$ but they may depend on all the rest of the coordinates.   In all such cases, the plane wave limit is the Minkowski spacetime.  Therefore {\sl all near horizon geometries of extreme Killing horizons admit Minkowski  spacetime as a plane wave limit}.  Though they may also admit other plane wave limits adapted to a different set of PCs.

\subsubsection{The $L=0$ congruence}

Next consider the kinematic regime for which $E, Q\not=0$ and $L=0$. The   HJ function is  $S=Ev+R(\rho)+Z(z)$, where
\bea
R(\rho)={Q\over \cos\theta} -\epsilon_\rho Q \tan\theta+\epsilon_\rho Q \theta~,~~~\rho={E\over Q} \cos\theta~,~~~\epsilon_\rho=\pm1~,
\eea
and
\bea
Z(z)=-\epsilon_z Q \psi~,~~~z=\cos\psi~,~~~\epsilon_z=\pm1~,
\eea
which follow from (\ref{nhorhj}).
In turn, the solution to the geodesic equation can be given as
\bea
&&{3\over2} \psi+{1\over4} \sin(2\psi)=-\epsilon_z Q \lambda+\psi_0~,~~~\theta=\epsilon_\rho \epsilon_z \psi+\theta_0~,~~~
\cr
&&v=-{Q\over E} \Big[\tan(-\epsilon_z \psi-\epsilon_\rho \theta_0)+\cos^{-1} (-\epsilon \psi-\epsilon_\rho \theta_0)\Big]+v_0~,
\cr
&&\chi=2 \Big[\log \tan\Big({-\epsilon_z \psi-\epsilon_\rho \theta_0\over 2}+{\pi\over4}\Big)-\log \cos (-\epsilon_z \psi-\epsilon_\rho \theta_0)\Big]+\chi_0~.
\label{geo2}
\eea
The first equation above can be used to express  $\psi$ in terms of the affine parameter $\lambda$.  Following the general description of the PCs for systems with separable HJ and geodesic equations, one writes
\bea
u=S=\epsilon_\rho Q \theta_0+E v_0~,~~~\lambda=\lambda~,~~~\phi=\chi_0~,~~~x=\theta_0~,
\eea
where $\psi_0$ is fixed, i.e. $\psi_0$ is set to zero.
The metric in these coordinates  reads as
\bea
&&g=du\Big[2 d\lambda +\epsilon_\rho {2\over Q} \Big((1+\cos^2\psi)+Q^2 Y_Q\cos^2(\epsilon_z\psi+\epsilon_\rho  x)\Big) dx
\cr
&&\quad
+{4\over Q}{\sin^2\psi\over 1+\cos^2\psi} \cos(\epsilon_z\psi +\epsilon_\rho x) d\phi +Y_Q\cos^2(\epsilon_z+\epsilon_\rho x) du   \Big]
\cr
&&\quad
+{\sin^2\psi\over 1+\cos^2\psi} \big(d\phi-2 \epsilon_\rho  \cos(\epsilon_z \psi+\epsilon_\rho x) dx\big)^2
\cr
&&
\quad+ (1+\cos^2\psi) \sin^2(\epsilon_z\psi+\epsilon_\rho x) dx^2~,
\label{lometr}
\eea
where $\psi=\psi(\lambda)$ is a solution of the first equation in (\ref{geo2}) and
\bea
Y_Q={3-6\cos^2\psi-\cos^4\psi\over Q^2(1+\cos^2\psi)}~.
\eea
The plane wave limit metric in Rosen coordinates becomes
\bea
g_{pw}=2 du d\lambda+{\sin^2\psi\over 1+\cos^2\psi} \big(d\phi-2 \epsilon_\rho  \cos( \psi) dx\big)^2+ (1+\cos^2\psi) \sin^2(\psi) dx^2~.
\eea
The plane wave metric can be written in $(u, \psi, x,\phi)$ coordinates using $(1+\cos^2\psi) d\psi=-\epsilon_z Q d\lambda$.

For small $\psi$, one has that $2\psi=-\epsilon_z Q \lambda$ and the plane wave limit is Minkowski spacetime.  For large $\lambda$, one has ${3\over2} \psi=-\epsilon_z Q \lambda$ and the metric (\ref{lometr}) can be explicitly expressed in terms of $\lambda$ as
\bea
ds^2&=&2 du d\lambda
+{\sin^2\big({2\over3} Q\lambda\big)\over 1+\cos^2\big({2\over3} Q\lambda\big)} \Big(d\phi-2 \epsilon_\rho  \cos\big({2\over3} Q\lambda\big) dx\Big)^2
\cr
&&
+ \Big(1+\cos^2\big({2\over3} Q\lambda\big)\Big) \sin^2\big({2\over3} Q\lambda\big) dx^2~,
\eea
in Rosen coordinates.

\subsubsection{General Case}

Next consider the general case with $E,Q,L\not=0$. The HJ equation (\ref{nhorhj}) for  $R$ can be solved to yield
%\bea
%\Big({dR\over d\rho}+{E\over \rho^2}-{2L\over \rho}\Big)^2=\Big({E\over\rho^2}-{2L\over \rho}\Big)^2-{Q^2\over \rho^2}~,
%\eea
%and the solution can be expressed as
\bea
R=2 L\log \rho+{E\over \rho}+\tilde R(\rho)~,
\eea
where
\bea
\tilde R(\rho)=\epsilon_\rho \int^\rho dp \sqrt{\Big({E\over p^2}-{2L\over p}\Big)^2-{Q^2\over p^2}}~.
\eea
Furthermore setting $z=\cos\psi$, one finds that
\bea
Z(z)=\Psi(\psi)=\epsilon_z \int^\psi d\beta\, {1\over \sin\beta} \sqrt{Q^2 \sin^2\beta-L^2 (1+\cos^2\beta)^2}~.
\eea
The geodesic equations can be expressed as
\bea
&&\dot v={1\over \cos^2\psi} {dR\over d\rho}~,~~~
\dot\psi={1\over 1+\cos^2\psi} {d\Psi\over d\psi}~,
\cr
&&
\dot\chi={1+\cos^2\psi\over \sin^2\psi} L-{2\rho\over 1+\cos^2\psi} {dR\over d\rho}~,
\cr
&&\dot\rho={E\over 1+\cos^2\psi}+ {\rho^2\over 1+\cos^2\psi} {dR\over d\rho}-{2 L \rho\over 1+\cos^2\psi}~.
\eea
The second equation above can be solved to express $\psi$ in terms of the affine parameter $\lambda$ as $\psi=\psi(\lambda)$. As in the previous case, we do not introduce a boundary condition for $\psi$. Then the last geodesic equation can be solved as $\rho=\rho(\lambda, \rho_0)$. Using this, one can solve the geodesic equation for $v$ and $\chi$ as  $v=V(\lambda, \rho_0)+v_0$ and $\chi={\sl X}(\lambda, \rho_0)+\chi_0$, respectively,
where
\bea
V=\int^\lambda d\mu {1\over 1+\cos^2\psi} {dR\over d\rho}~,~~~{\sl X}=\int^\lambda d\mu \Big({1+\cos^2\psi\over \sin^2\psi} L-{2\rho\over 1+\cos^2\psi} {dR\over d\rho}\Big)~.
\label{inta}
\eea
To express the metric in PCs, set $x=\rho_0$, $\phi=\chi_0$,  $\lambda=\lambda$ and $u=S(\rho_0,\chi_0, v_0)$.  Eliminating $v_0$ in favour of the $u$ coordinate, one finds that the metric in the new coordinates can be expressed as
\bea
&&\Delta={3-6\cos^2\phi-\cos^4\psi\over E^2(1+\cos^2\psi)}~,~~~h_\phi=-L \lambda\, \Delta+{2\sin^2\psi\over E(1+\cos^2\psi)}
\cr
&&
h_x={2\sin^2\psi\over E(1+\cos^2\psi)} \partial_{\rho_0}{\sl X}+ {1\over E\lambda\,} (1+\cos^2\psi) \partial_{\rho_0}\rho-\Delta \lambda\, \partial_{\rho_0}R(\rho_0) +E\Delta \lambda\,
\partial_{\rho_0} V~,
\cr
&&
\gamma_{\phi\phi}={\sin^2\psi\over 1+\cos^2\psi}-{4\lambda\,L\sin^2\psi\over E(1+\cos^2\psi)}+L^2 \Delta \lambda^2~,
\cr
&&
\gamma_{xx}=E^2 \Delta (E^{-1} \partial_{\rho_0} R(\rho)-\partial_{\rho_0} V)^2-2\lambda\,{\sin^2\psi\over 1+\cos^2\psi}\partial_{\rho_0}{\sl X}(E^{-1} \partial_{\rho_0} R(\rho)-\partial_{\rho_0} V)
\cr
&&
\qquad-2 \partial_{\rho_0} \rho (1+\cos^2\psi)(E^{-1} \partial_{\rho_0} R(\rho)-\partial_{\rho_0} V)+{\sin^2\psi\over 1+\cos^2\psi}(\partial_{\rho_0}{\sl X})^2~,
\cr
&&
\gamma_{x\phi}=-{L\over E} (1+\cos^2\psi) \partial_{\rho_0} \rho+E^2\Delta \lambda^2 (E^{-1} \partial_{\rho_0} R(\rho)-\partial_{\rho_0} V)
\cr
&&\qquad
+{\sin^2\psi\over 1+\cos^2\psi} \Big((1-2\lambda\,L E^{-1})\partial_{\rho_0}{\sl X}+2\lambda\, \partial_{\rho_0}V-2\lambda\, E^{-1} \partial_{\rho_0} R(\rho_0)\Big)~.
\eea
The plane wave limit of this metric can be easily taken and it will  not stated here. The above expression of the metric can become explicit provided one can carry out the integrals in (\ref{inta}) and so  integrate the null geodesic equation.  It may not be possible to express such integrals in terms of ``simple'' functions, see for example \cite{gralla} for related studies.

\section{Concluding Remarks}

We have demonstrated that for gravitational backgrounds for which both null HJ and geodesic equations separate, there is a systematic way to express the metric in  PCs and continue to take the plane wave limit of these spacetimes. Although the above separability properties of the spacetime allows one to give a close expression of the metric in PCs and derive  plane wave limits of these spacetimes, it does not necessary lead to explicit expressions for the metric in PCs and its plane wave limit as the latter requires  an  explicit solution of a null geodesic in the separation coordinates.  Such a solution may not be written in terms of ``simple'' functions leading to rather involved expressions for the metric for generic choices of a geodesic.  Nevertheless some plane wave limits have been found  and it has been emphasized that the near horizon geometries of all extreme regular Killing horizons admit Minkowski spacetime as a plane wave limit.  The  propagation of strings in these plane wave limits will be presented elsewhere.

\section*{Acknowledgments}

I would like to thank Dionysis Anninos for bringing \cite{gralla} into my attention.

%https://en.wikipedia.org/wiki/List_of_integrals_of_irrational_functions

\end{document}